\documentclass[final,1p,times]{elsarticle} 
\usepackage{graphicx} 
\usepackage{amssymb} 
\usepackage{amsthm} 
\usepackage{lineno} 

\journal{Nuclear Physics A} 
\begin{document} 

\begin{frontmatter} 


  \title{J/$\psi$ Elliptic Flow, High $p_T$ Suppression and $\Upsilon$
    Measurements in A+A Collisions by the PHENIX Experiment.}

\author{Ermias T. ATOMSSA for the PHENIX Collaboration}

\address{Stony Brook University, Department of Physics and Astronomy,
  Stony Brook, NY, 11794, USA}

\begin{abstract} 
  Three measurements that broaden the scope of the experimental
  investigation of quarkonia modifications in heavy ion collisions are
  presented. Although the current statistical precision on the first
  two measurements does not allow one to draw significant conclusions,
  J/$\psi$ elliptic flow and high $p_T$ suppression results are
  important proofs of principle measurements, and make the case for
  higher luminosities at RHIC. Finally, the first measurement of an
  upper limit on the nuclear modification $R_{AA}$ of dielectrons in
  the $\Upsilon$ mass region in Au+Au collisions is presented. The
  results show a significant suppression with an upper limit of
  $<$~0.64 at 90\% CL.
\end{abstract} 

\end{frontmatter} 




\section{Introduction}\label{sec:intro}

J/$\psi$ suppression measurements in heavy ion collisions have long
been used as a test of deconfinement in the Quark Gluon Plasma (QGP)
formed in heavy ion collisions. J/$\psi$ rates in heavy ion collisions
are now known to be subject to a number of competing mechanisms of
production and suppression. These mechanisms can be categorized into
Cold Nuclear Matter (CNM) such as breakup through nuclear interaction
and shadowing and QGP induced mechanisms including the looked for
melting induced by Debye screening, and recombination.

The use of J/$\psi$ to study dissociation phenomenon due to Debye
screening in the QGP requires the understanding and good control of
every other mechanism. It is therefore essential to broaden the scope
of experimental observables. Here the focus will be on three
particularly challenging measurements, where results are in their
beginning stages at RHIC. The critically important measurement of CNM
effects is not covered, as it was the subject of another talk at this
conference~\cite{csilva,mleitch}.

\section{Momentum dependence of J/$\psi$ suppression}\label{sec:hipt}

\begin{figure}[hbpt]
  \centering
  \includegraphics[width=0.44\columnwidth]{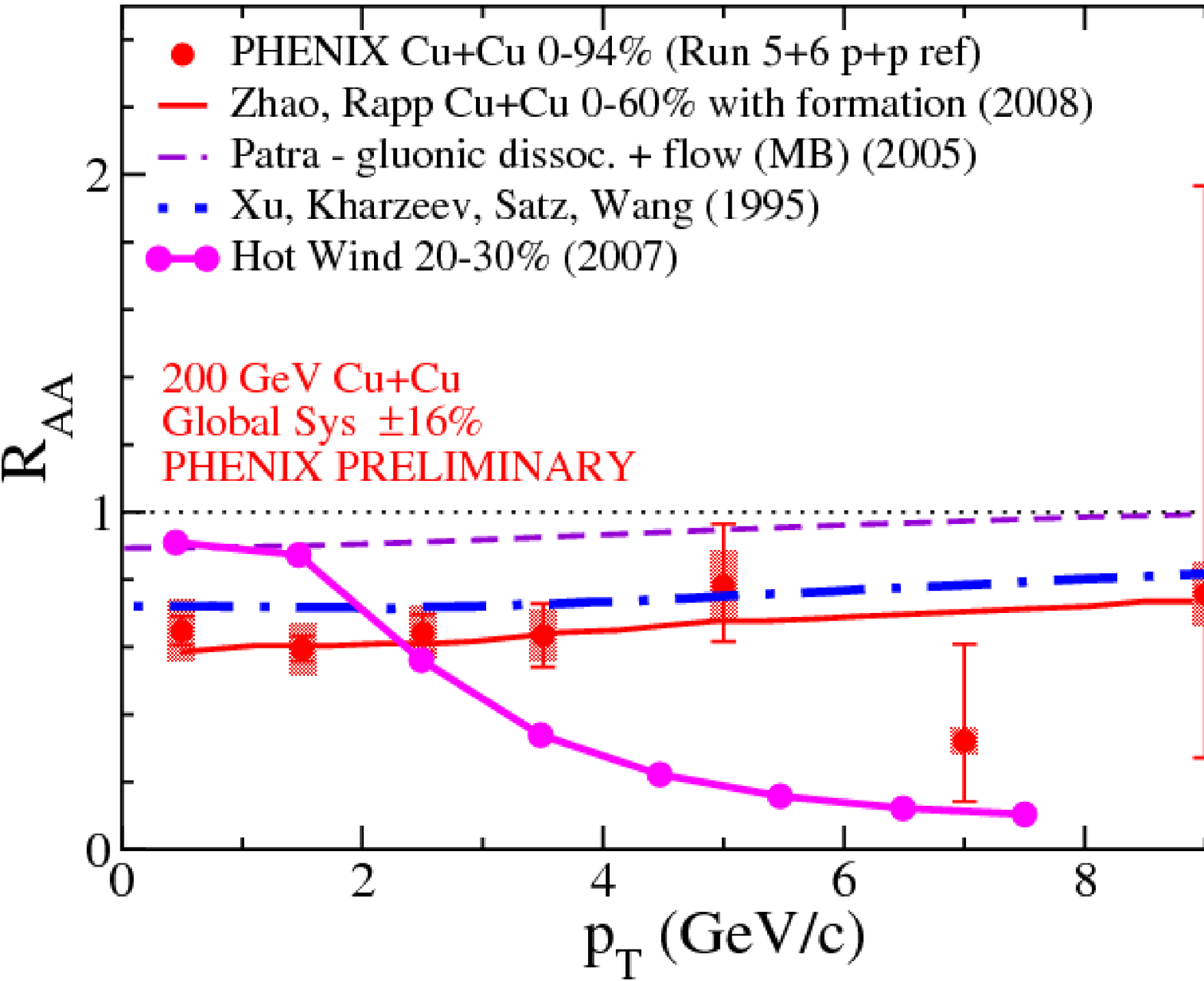}
  \includegraphics[width=0.50\columnwidth]{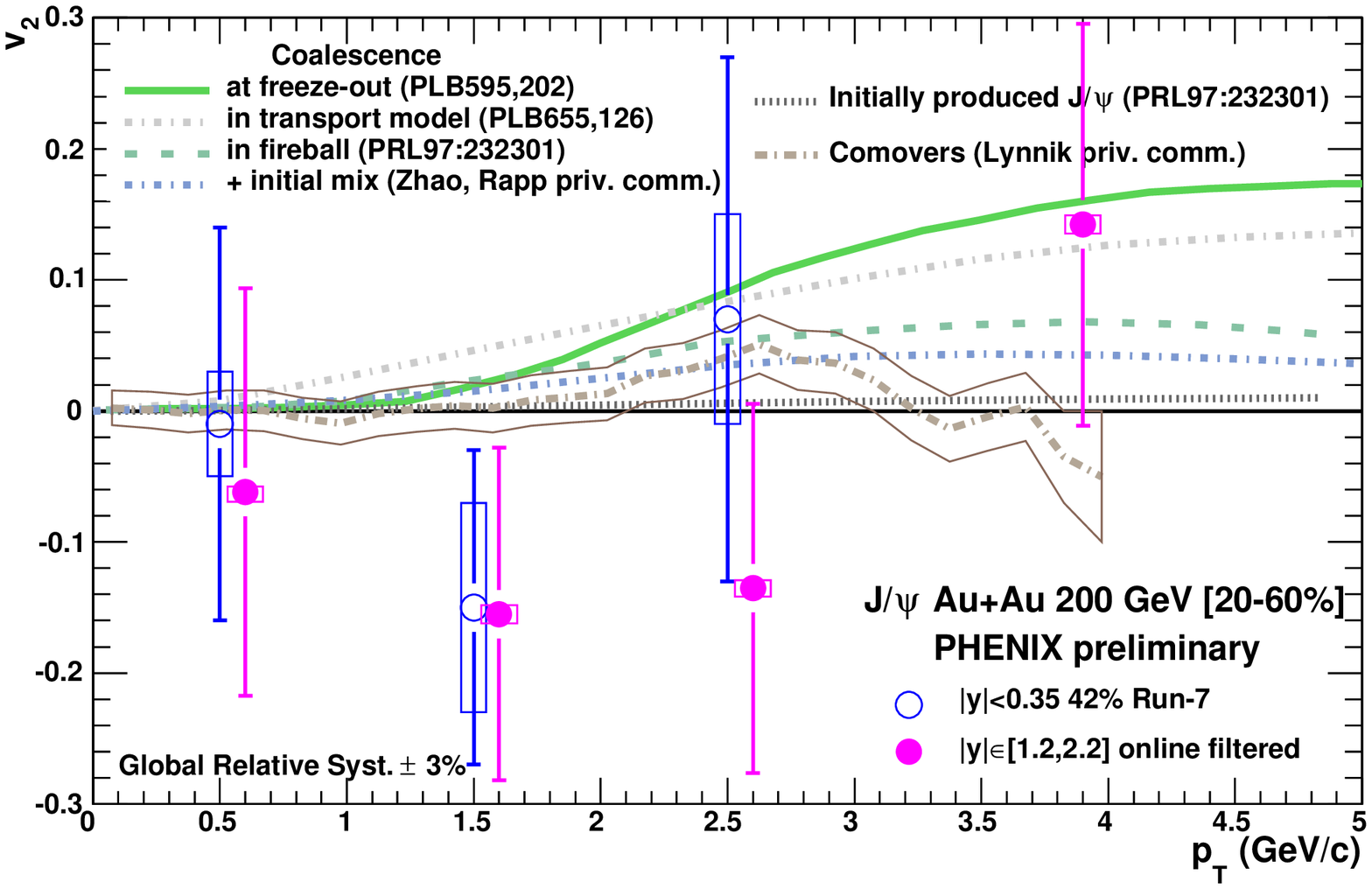}
  \caption{Left: J/$\psi$ suppression in 0-20\% central Cu+Cu
    collisions vs. $p_T$ at mid ($|y|<0.35$) rapidity region. Right:
    Elliptic flow of J/$\psi$ vs. $p_T$ in the PHENIX mid and forward
    ($1.2<|y|<2.2$) rapidity regions. }
  \label{fig:flow_hipt}
\end{figure}

The momentum dependence of J/$\psi$ suppression is a very useful
indication of some of the initial and final state effects. On the one
hand, initial state multiple scattering which is present in
proton/deuteron--ion and ion--ion collisions is expected to broaden
the $p_T$ spectrum with respect to the reference proton--proton
spectra, due to a random walk of the colliding partons in the
transverse plane to the collision axis thorough elastic scattering
before the final inelastic interaction that creates the $c\bar{c}$
state. This effect is known as the Cronin effect. On the other hand,
recombination, if it takes place in the QGP is expected to result in a
softer $p_T$ spectrum than in proton--proton collisions. This comes
from the expectation that only $c$ and $\bar{c}$ pairs close in
momentum space actually combine, and that the low $p_T$ charm quarks
dominate the spectrum. Measuring J/$\psi$ suppression at the highest
possible $p_T$ is essential since different model predictions show the
most variation in this region. Figure~\ref{fig:flow_hipt} (left) shows
the currently available J/$\psi$ $R_{AA}$ versus $p_T$ in 0-20\%
central Cu+Cu collisions. The data clearly disfavors the hot wind
model~\cite{hotwind}, whereas more statistical precision is required
to discriminate between the other models~\cite{twocomp}.

\section{J/$\psi$ elliptic flow}\label{sec:flow}

The physics motivation for the measurement of the J/$\psi$ elliptic
flow comes from the observation that semileptonic decay electrons from
heavy flavor show a strong flow that is thought to be accounted for by
a strong thermalization of the underlying heavy
quarks~\cite{openc}. As a result, J/$\psi$ created by recombination of
independent charm quarks will inherit the flow of their constituents,
whereas direct J/$\psi$ possess at most a very small $v_2$ due to
geometrical effects in their absorption. The measurement of J/$\psi$
elliptic flow has thus the potential to discriminate between different
recombination scenarios, as illustrated by the theoretical curves
shown in Figure~\ref{fig:flow_hipt} (right), predicted by models that
incorporate various degrees of
regeneration~\cite{regen}. Unfortunately the current measurement of
J/$\psi$ $v_2$ from PHENIX at mid and forward rapidities, shown in the
same figure, are limited by statistics and do not favor any particular
scenario.

\section{Insight from higher mass states}\label{sec:ups}

In this section, the efforts to measure the nuclear modification of
$b\bar{b}$ resonances are described. The $b\bar{b}$ states are more
tightly bound than the $c\bar{c}$ states. As a result most lattice QCD
calculations predict a much higher melting temperature for the lowest
$\Upsilon$ state than for the J/$\psi$, despite discrepancies on the
actual value of the melting temperatures~\cite{lqcd}. The tighter
binding also implies that $b\bar{b}$ state should suffer less CNM
breakup. Pinning down the $\Upsilon$ nuclear modification factor
$R_{AA}$ will provide new constraints in the study of quarkonium
modifications.

\begin{figure}[hbpt]
  \centering
  \includegraphics[width=0.44\columnwidth]{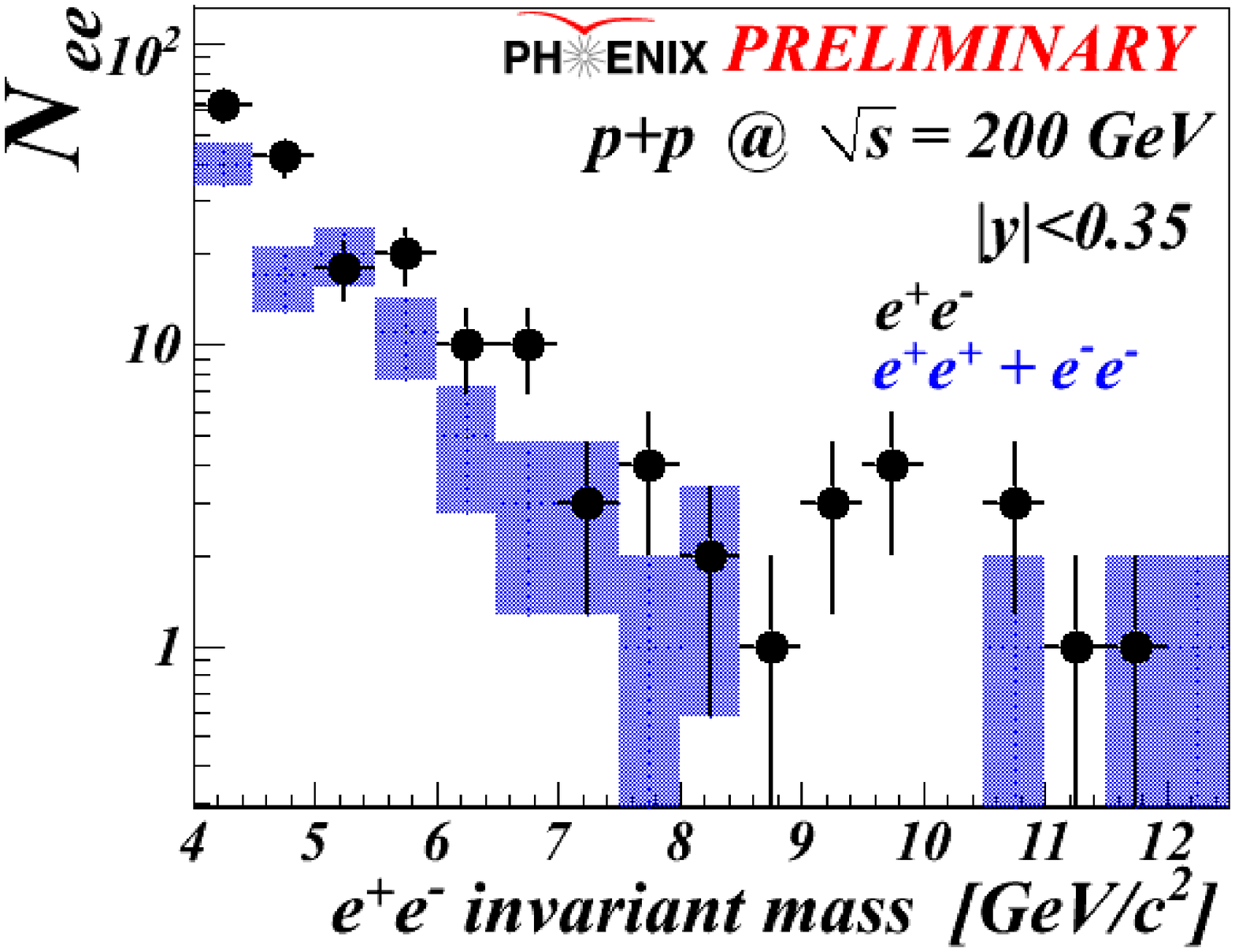}
  \includegraphics[width=0.54\columnwidth]{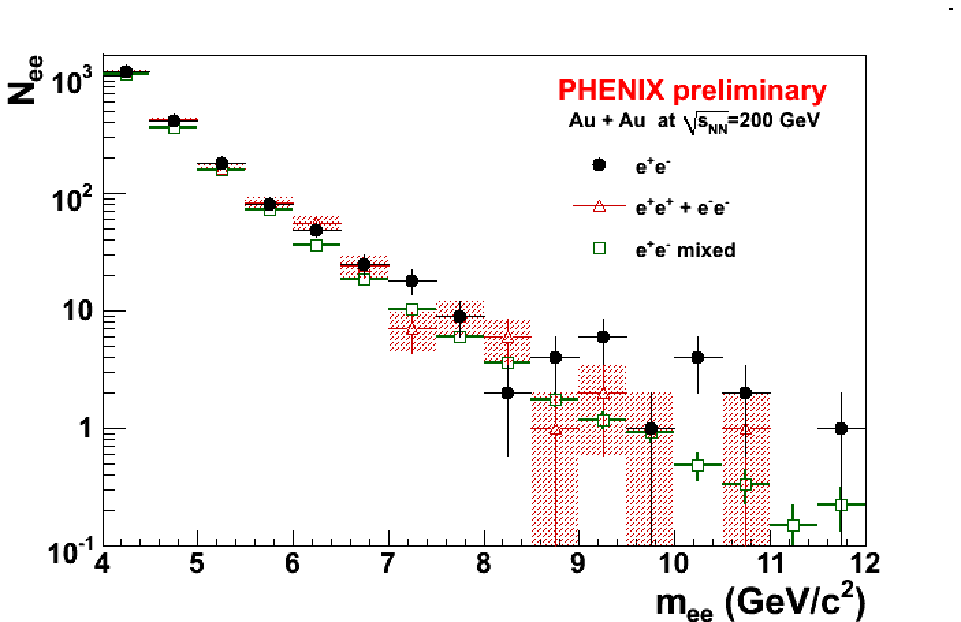}
  \caption{Invariant mass spectra in p+p (left) and Au+Au
    (right). Closed circles are used for the unlike sign spectrum,
    whereas the like sign background spectrum and its error are plotted
    as shaded bands. On the right hand plot, the mixed event unlike
    sign background spectrum is also plotted for comparison as open
    squares.}
  \label{fig:inv_mass_upsilon}
\end{figure}

The measurement of $\Upsilon$ is however more challenging at RHIC than
J/$\psi$ due to the smaller bottom production cross section, and as a
result requires much higher integrated luminosity to attain the same
level of statistical significance as a J/$\psi$
measurement. Figure~\ref{fig:inv_mass_upsilon} shows the foreground
and combinatorial background invariant mass spectra in the high mass
region in p+p (left) and Au+Au (right) interactions and corresponding
combinatorial background. Most of the available data from the 2006 p+p
RHIC run and 2007 Au+Au RHIC run at 200~GeV were included here. The
net like and unlike pair yields in p+p and Au+Au in the mass region
from 8.5~GeV/$c^2$ to 11.5~GeV/c$^2$ are summarized in
Table~\ref{tab:counts}.

\begin{table}[hbpt]
  \centering
  \begin{tabular}{{|c|c|c|c|}}
    \hline
    &  Unlike sign   &  Like sign &  \\
    &  [8.5-11.5]~GeV/c$^2$  &  [8.5-11.5]~GeV/c$^2$ & J/$\psi$ yield \\
    \hline
    \hline
    p+p    &  12 & 1  & 2653 $\pm$ 70 (stat) $\pm$ 345 (syst)  \\
    \hline
    Au+Au   &  17  &  5  &   4166 $\pm$ 442 (stat) $\pm^{187}_{304}$ (syst)\\
    \hline
  \end{tabular}
  \caption{Net J/$\psi$ yield and like and unlike sign pair yields in
    the [8.5-11.5]~GeV/c$^2$ mass range for p+p and Au+Au collisions.}
  \label{tab:counts}
\end{table}

\begin{figure}[hbpt]
  \centering
  \includegraphics[width=0.7\columnwidth]{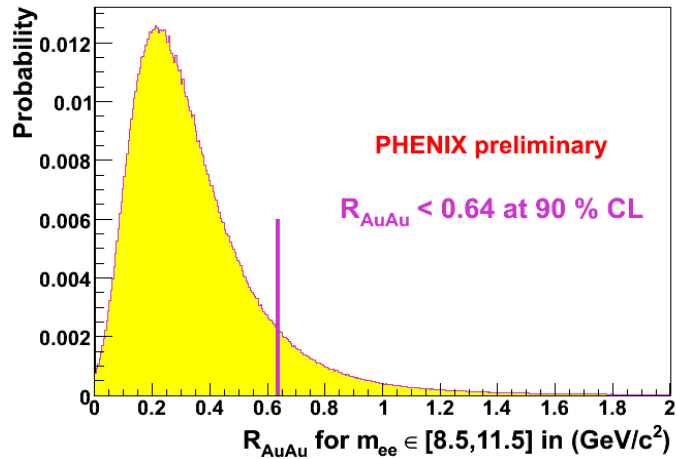}
  \caption{The statistical probability distribution of the $R_{AA}$ of
    the signal from correlated pairs as obtained from
    Eq.~\ref{eq:raa}. An upper limit at 90\% confidence level of 0.64
    is inferred from this distribution.}
  \label{fig:upsilon_suppression}
\end{figure}

In order to take advantage of canceling systematic uncertainties, the
$R_{AA}$ in the $\Upsilon$ mass region was calculated by taking its
ratio to that of the J/$\psi$,
$R_{AA}=0.42\pm0.025(stat)\pm0.051$~\cite{auau}.

\begin{equation}
  \frac{R_{AA}([8.5,11.5])}{R_{AA}(J/\psi)} = 
  \big(
  \frac{N([8.5,11.5])_{AA}/A\epsilon(\Upsilon)_{AA}}{N_{coll} \times
    N([8.5,11.5])_{pp}/A\epsilon(\Upsilon)_{pp}}
  \big)/
  \big(
  \frac{N(J/\psi)_{AA}/A\epsilon(J/\psi)_{AA}} {N_{coll} \times
    N(J/\psi)_{pp}/A\epsilon(J/\psi)_{pp}}
  \big)
\nonumber
\end{equation}

and rearranging to get 

\begin{equation}
R_{AA}([8.5,11.5]) =
\frac{(N[8.5,11.5]/N(J/\psi))_{AA}}{(N[8.5,11.5]/N(J/\psi))_{pp}}
\times R_{AA}(J/\psi) \nonumber
\label{eq:raa}
\end{equation}

where the acceptance and efficiencies cancel out in the double
ratio. This approach has the advantage that efficiency calculation
systematics cancel in the ratio $N[8.5,11.5]/N(J/\psi)$ within a given
data set. The probability distribution for $R_{AA}([8.5,11.5])$ shown
in Figure~\ref{fig:upsilon_suppression} is calculated by a convolution
of the Poissonian probability distributions for
$N[8.5,11.5]/N(J/\psi)$ in Au+Au and p+p deduced from the values in
Table~\ref{tab:counts} and smeared by the Gaussian distributed error
of $R_{AA}(J/\psi)$. From this distribution, the upper limit on
$R_{AA}([8.5,11.5])$ at 90\% confidence level is determined to be
64\%~\cite{zconesa}.

The following considerations have to be taken into account when trying
to interpret this value. The first is that the contributions of other
correlated physics backgrounds like open beauty and Drell-Yan are
estimated to be quite small, of the order of 15\% or less from an
extrapolation of the lower mass spectrum in p+p. Although physics
background subtraction was not performed before calculating the upper
limit given here, its contribution seems to be small. The other
consideration is that the high mass counts integrate a mass range that
was determined as the experimental width covered by the
$\Upsilon$(1S), $\Upsilon$(2S) and $\Upsilon$(3S) states. It also
includes a feed down contribution from excited states such as the
$\chi_b$. The suppression observed is therefore a cumulative effect of
the above mentioned resonances, which, except for the 1S ground state
are predicted by many lattice QCD calculations to have a melting
temperature close to that of J/$\psi$s. Finally, any interpretation of
this result should take into consideration the cold nuclear matter
effects such as break up and shadowing on the high mass resonances.


\end{document}